\documentclass[reprint, 10pt, a4paper, aps, prx, floatfix, longbibliography]{revtex4-1}
\usepackage[utf8]{inputenc}
\usepackage{amsmath}
\usepackage{amsfonts}
\usepackage{amssymb}
\usepackage{mathrsfs}
\usepackage{subfigure}
\usepackage[hidelinks]{hyperref}
\usepackage{outlines}
\usepackage{tikz}
\usepackage{booktabs}

\usetikzlibrary{decorations.pathmorphing}

\newcommand{\bv}{\mathbf{v}}
\newcommand{\bE}{\mathbf{E}}
\newcommand{\bB}{\mathbf{B}}

\newcommand{\D}{\mathrm{d}}

\begin{document}

\title{Natural Hot Ion Modes in a Rotating Plasma}
\date{\today}
\author{E. J. Kolmes}
\email[Electronic mail: ]{ekolmes@princeton.edu}
\author{I. E. Ochs}
\author{M. E. Mlodik}
\author{N. J. Fisch}
\affiliation{Department of Astrophysical Sciences, Princeton University, Princeton, New Jersey 08540, USA}

\begin{abstract}

In steady state, the fuel cycle of a fusion plasma requires inward particle fluxes of fuel ions. 
These particle flows are also accompanied by heating. 
In the case of classical transport in a rotating cylindrical plasma, this heating can proceed through several distinct channels depending on the physical mechanisms involved. 
Some channels directly heat the fuel ions themselves, whereas others heat electrons. 
Which channel dominates depends, in general, on the details of the temperature, density, and rotation profiles of the plasma constituents. 
However, remarkably, under relatively few assumptions concerning these profiles, if the alpha particles, the byproducts of the fusion reaction, can be removed directly by other means, a hot-ion mode tends to emerge naturally. 

\end{abstract}
	
\maketitle
	
\section{Introduction}

Nuclear fusion devices aim to achieve ignition by heating a plasma to a very high temperature, typically on the order of tens of keV.  The heat losses at these temperatures are a significant source of inefficiency in a fusion device. However, the fusion cross-section depends only on the temperature of the fuel ions.  At the same time, hot electrons incur large power losses, either through radiation or heat transport, but do not produce fusion power. Moreover, the capacity of a magnetic confinement device to trap plasma is typically limited by total plasma pressure; thus, the higher-temperature electrons take up a large share of that pressure limit without producing any additional fusion power. As such, the performance of a fusion device can be improved -- often dramatically -- by achieving a ``hot-ion mode," in which the ions are maintained at a higher temperature than the electrons \cite{Clarke1980,Fisch1994}. 

However, attaining a hot-ion mode is a significant technical challenge. High-energy ions produced by fusion preferentially lose their energy collisionally to electrons rather than to fuel ions. If no additional strategy is employed to heat the ion population, the electrons will tend to be at least as hot as the fuel ions, if not hotter. A hot-ion mode can be produced if significant external heating sources are directed at the ion population. These sources could be neutral beams or RF waves. 
Attaining a hot-ion mode in a reactor, however, where the main heating is necessarily through the fusion reaction, requires some form of $\alpha$-channeling, in which the energy from fusion byproducts is channeled into a wave (avoiding collisional heating of the electrons), and that wave deposits its energy into the fuel ions \cite{Fisch1992,Valeo1994,Fisch1995ii,Fisch1995,Ochs2015b,Cianfrani2018,Cianfrani2019,Castaldo2019,Romanelli2020}. In all of these cases, the hot-ion mode requires significant intervention to change the power balance such that energy is directed to fuel ions. In any of these cases, the differential in temperatures could be increased if the electron energy confinement were reduced, though this strategy is less desirable insofar as it involves increasing energy losses. 

This paper will suggest an alternative possibility: a ``natural" hot-ion mode.    The notion of ``natural" requires definition.  By {\it natural}, we imagine processes  in which the ion heating comes from transport processes that are already happening in the plasma.   Note that any steady-state fusion device must have inward flows of fuel ions and outward flows of alpha particles in order to balance the fusion reactions. In the case of classical transport, each particle flux is accompanied by dissipation. If the dissipation is directed into the ions, and if it is sufficiently large, then it could be possible to reach a hot ion mode {\it naturally}, without having to heat the ions externally. 

A plasma with large electric fields is a logical place to look for such an effect, since the electric fields provide a reservoir of potential energy through which moving particles can exchange energy. 
If there are transfers of energy between the particles and the electrostatic potential, one might imagine that certain populations of particles could be preferentially heated or cooled. 
Exploiting this possibility, this paper will consider cylindrically symmetric configurations with radial electric and axial magnetic fields. We will focus on a particularly simple case, in which the transport is purely classical and where inhomogeneities in the direction of the field can be neglected. In other words, consider the cylinder long enough for end losses to be less important than transport across the field. 

Perpendicular $\bE$ and $\bB$ fields in this geometry cause plasmas to rotate, so any discussion of crossed-field plasmas is inevitably a discussion of rotating plasmas. 
There are a number of proposals for fusion devices involving significant rotation. 
Supersonically rotating magnetic mirror devices have promising theoretical properties \cite{Lehnert1971, Bekhtenev1980} and have been realized experimentally \cite{Ellis2001, Ellis2005, Teodorescu2010}. 
The wave-driven rotating torus (or WDRT) is a proposal for a toroidal device which relies on poloidal rotation for confinement \cite{Rax2017, Ochs2017ii}. 
Additionally, significant rotation velocities are sometimes observed even in devices (like tokamaks) for which rotation is not an essential part of the confinement scheme \cite{Suckewer1979, Kim1991, Conway2004, HelanderSigmar, Rice2007, deGrassie2009, StoltzfusDueck2012}. 

The dissipation due to classical cross-field transport in a rotating plasma can be split into two categories. 
Viscous heating occurs when the rotation profile is sheared. It heats ions. 
Frictional heating occurs due to interactions between particles of different species with different velocities. 
Frictional interactions between ions and electrons heat electrons. 
However, in the case of a plasma containing more than one ion species, there is also ion-ion frictional dissipation, in which the heat is divided between the ion species, preferentially heating the lighter of any pair. 

The problem of controlling the relative importance of these channels has largely been overlooked, despite the fact that expressions for the heating follow readily from established theories of classical transport. 
The viscous heating scales with the size of the deviation of the rotation profile from solid-body rotation; the frictional heating scales with the deviation of the different species' pressures from a class of dissipationless profiles. 

We show here  that a hot ion mode can arise naturally by arranging for the dominance of the ion dissipation channel. 
For a plasma with a single ion species, classical dissipation directly heats the ions only if the viscous dissipation is large. 
If there are multiple ion species, that constraint is relaxed, since ion-ion friction can then compete with ion-electron friction. 
In general, in order for the heat dissipated in the rotating plasma to naturally flow to the ions,  the temperature, density, and rotation profiles of the plasma constituents must all be arranged carefully.  

However, we further show here a remarkable property of rotating and fusing plasma, so long as the steady state density of ions is maintained through ion fueling balancing the prompt removal by $\alpha$-channeling of the spent fusion byproducts. 
Any other mechanism that removes the fusion byproducts on a collisionless timescale would produce the same result. 
Surprisingly, such a steady-state plasma tends to assume naturally the very favorable hot-ion mode without the necessity of arranging in detail these profiles. 

This paper is organized as follows: Section~\ref{sec:particleTransport} reviews classical cross-field particle transport in a rotating plasma. 
Section~\ref{sec:classicalHeating} describes the classical heating channels  associated with the different mechanisms of cross-field particle transport. 
Section~\ref{sec:control} discusses the conditions allowing different channels to be dominant. 
Section~\ref{sec:globalConditions} describes how, specifying only global particle balance and boundary conditions, the ion channel can dominate. 
Section~\ref{sec:examples} briefly discusses two types of device in which dominant ion heating could be particularly dramatic. 
Section~\ref{sec:realizability} enumerates and discusses the major assumptions that this paper relies on. 
Section~\ref{sec:discussion} summarizes and discusses these results. 

\section{Classical Particle Transport in a Rotating Linear Device} \label{sec:particleTransport}

Consider a fully ionized cylindrical plasma device with an axial field $\bB = B \hat z$. Suppose the system is homogeneous in the $\hat \theta$ and $\hat z$ directions, and that all flows are radial or azimuthal. In a region away from any particle sources or sinks, the momentum equation for species $s$ is 
\begin{align}
&m_s n_s \bigg( \frac{\partial \bv_s}{\partial t} + \bv_s \cdot \nabla \bv_s \bigg) \nonumber \\
&\hspace{25 pt}= q_s n_s (\bE + \bv_s \times \bB) - \nabla p_s - \nabla \cdot \pi_s + \mathbf{R}_s, \label{eqn:momentum}
\end{align}
where $m_s$ is the mass of species $s$; $n_s$ is the density; $\bv_s$ is the velocity; $q_s$ is the charge; $p_s$ is the pressure; $\nabla \cdot \pi_s$ is the viscous force density; $\mathbf{R}_s$ is the friction force density; and $\bE$ is the electric field. In steady state, the radial flux $\Gamma_s \doteq n_s v_{sr}$ can be obtained \cite{Kolmes2019} by rearranging the $\hat \theta$ component of Eq.~(\ref{eqn:momentum}): 
\begin{gather}
\Gamma_{s} = \frac{R_{s \theta} - (\nabla \cdot \pi_s)_\theta}{m_s \Omega_s [1 + (r v_{s \theta})' / r \Omega_s]} \, . \label{eqn:radialFlux}
\end{gather}
Here the $\theta$ subscript denotes the $\hat \theta$ component of a vector, and $\Omega_s \doteq q_s B / m_s$ is the gyrofrequency. 
The prime in the denominator denotes the derivative $\partial / \partial r$. 
The friction force density can be written as 
\begin{align}
R_{s \theta} = \sum_{s'} R_{ss' \theta},
\end{align}
where the $\hat \theta$ frictional force on species $s$ due to interactions with species $s'$ can be modeled by \cite{Hirshman1981}
\begin{align}
&R_{s s' \theta} = n_s m_s \nu_{ss'} \bigg[ (v_{s' \theta} - v_{s \theta}) \nonumber \\
&\hspace{12 pt}+ \frac{3}{2 B} \frac{1}{m_s T_{s'} + m_{s'} T_s} \bigg( \frac{m_{s'} T_s T_s'}{q_s} - \frac{m_s T_{s'} T_{s'}'}{q_{s'}} \bigg) \bigg] . \label{eqn:explicitFriction}
\end{align}
$T_s$ denotes the temperature of species $s$. $\nu_{ss'}$ is the collision frequency between species $s$ and $s'$. In order to ensure momentum conservation, it must satisfy $n_s m_s \nu_{ss'} = n_{s'} m_{s'} \nu_{s's}$, so that $R_{ss'\theta} + R_{s's\theta} = 0$. Eq.~(\ref{eqn:explicitFriction}) includes both the friction due to differences in flow velocity and the thermal friction, which is driven by temperature gradients. The azimuthal friction force that appears in Eq.~(\ref{eqn:radialFlux}) can be written as 
\begin{gather}
(\nabla \cdot \pi)_\theta = - \frac{1}{r^2} \frac{\partial}{\partial r} \bigg[ \eta_{s1} r^3 \frac{\partial}{\partial r} \bigg( \frac{v_{s \theta}}{r} \bigg) \bigg], 
\end{gather}
where $\eta_{s1}$ is the corresponding Braginskii viscosity coefficient \cite{Braginskii1965}, or its analog in a multiple-ion-species plasma \cite{Zhdanov}. 
This form of the viscous force is derived in Appendix~\ref{appendix:viscosity}. 

\begin{figure}
\centering
\includegraphics[width=\linewidth]{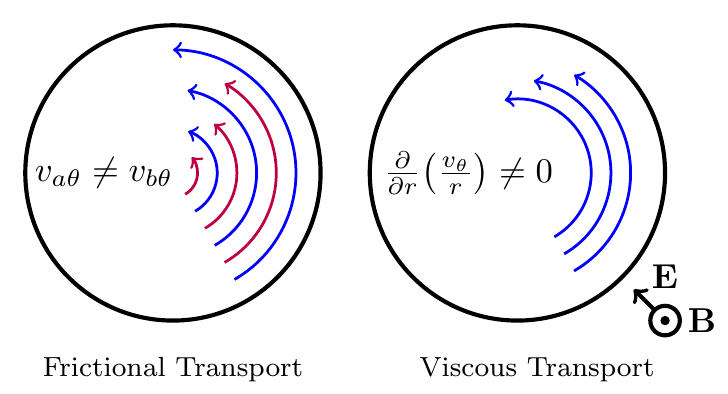}
\caption{The cartoon on the left shows the kind of flow that gives rise to frictional cross-field transport: the local velocities of different species are not the same. The cartoon on the right shows the kind of flow that gives rise to viscous cross-field transport: it is radially sheared. }
\label{fig:flowCartoon}
\end{figure}

Let $\delta \doteq v_{s \theta} / r \Omega_s$, and suppose $\delta \ll 1$. Then 
\begin{align}
\Gamma_{s} &= \frac{R_{s \theta} - (\nabla \cdot \pi_s)_\theta}{m_s \Omega_s} \, \sum_{k=0}^\infty \bigg[ - \frac{(r v_{s \theta})'}{r \Omega_s} \bigg]^k \\
&= \frac{R_{s \theta} - (\nabla \cdot \pi_s)_\theta}{m_s \Omega_s} \, \bigg[ 1 + \mathcal{O}(\delta) \bigg]. \label{eqn:fluxWithCorrection}
\end{align}
Define the viscous flux $\Gamma_s^\text{visc}$ by 
\begin{gather}
\Gamma_{s}^\text{visc} \doteq - \frac{(\nabla \cdot \pi_s)_\theta}{m_s \Omega_s} \, .
\end{gather}
Define the frictional flux $\Gamma_s^\text{fric}$ by 
\begin{gather}
\Gamma_s^\text{fric} \doteq \sum_{s'} \Gamma_{ss'}^\text{fric} \\
\Gamma_{ss'}^\text{fric} \doteq \frac{R_{ss'\theta}}{m_s \Omega_s} \, .
\end{gather}
Then Eq.~(\ref{eqn:fluxWithCorrection}) can be rewritten as 
\begin{gather}
\Gamma_s = ( \Gamma_s^\text{fric} + \Gamma_s^\text{visc} ) \big[ 1 + \mathcal{O}(\delta) \big]. 
\end{gather}
The kinds of flows that give rise to these different fluxes are shown in Figure~\ref{fig:flowCartoon}. 
In practice, the viscous flux $\Gamma_s^\text{visc}$ is often small compared to the frictional flux $\Gamma_{s}^\text{fric}$ \cite{Kolmes2019}. However, for the purposes of understanding heat and charge transport, the two fluxes are comparably important. 
This is partly because the flux described by $\Gamma_{ss'}^\text{fric}$ is ambipolar, in the sense that 
\begin{gather}
q_s \Gamma_{ss'}^\text{fric} + q_{s'} \Gamma_{s's}^\text{fric} = 0.
\end{gather}
The viscous flux $\Gamma_s^\text{visc}$ satisfies no such condition. 
In the discussions that follow, the ability of the flux to carry net charge will be important, since it determines how effectively the particles can exchange energy with an electrostatic potential. 

\section{Classical Heating} \label{sec:classicalHeating}

The particle fluxes described by Eq.~(\ref{eqn:radialFlux}) are accompanied by dissipation. 
The temperature evolution equation for species $s$ can be written as
\begin{align}
&\frac{3 n_s}{2} \bigg( \frac{\partial T_s}{\partial t} + \bv_s \cdot \nabla T_s \bigg) + \nabla \cdot \mathbf{q}_s + p_s \nabla \cdot \bv_s \nonumber \\
&\hspace{28 pt}= \sum_{s'}\frac{3 m_s n_s \nu_{ss'}}{m_s + m_s'} \big(T_{s'} - T_s \big) + Q_s^\text{visc} + Q_s^\text{fric} , \label{eqn:heat}
\end{align}
were $\mathbf{q}_s$ is the heat flux, $Q_s^\text{visc}$ is the viscous heating, and $Q_s^\text{fric}$ is the frictional heating. 
$Q_s^\text{visc}$ and $Q_s^\text{fric}$ are not the only terms in Eq.~(\ref{eqn:heat}) that could possibly produce a difference between different species' temperatures. 
The heat flows can have a significant impact on the evolution of $T_s$, as can compressional heating. 
The compressional heating is discussed in more detail in Appendix~\ref{appendix:compressionalHeating}. 
However, the goal of this paper is not to describe full solutions to Eq.~(\ref{eqn:heat}). 
Rather, our focus will be the ways in which $Q_s^\text{visc}$ and $Q_s^\text{fric}$ may preferentially heat different species. 
These are the two terms that are driven directly by classical collisional transport; each has a connection to one of the particle fluxes described in Section~\ref{sec:particleTransport}. 

\subsection{Viscous Heating}

The heating $Q_s^\text{visc}$ due to the viscous flux $\Gamma_{sr}^\text{visc}$ is the simpler of the two to understand. In this geometry, the leading-order viscous heating for species $s$ can be written as 
\begin{align}
Q_s^\text{visc} = - \pi_s : \nabla \bv_s = \eta_{s1} \bigg[ r \frac{\partial}{\partial r} \bigg( \frac{v_{s \theta}}{r} \bigg) \bigg]^2. \label{eqn:viscousHeating}
\end{align}
Eq.~(\ref{eqn:viscousHeating}) is derived in Appendix~\ref{appendix:viscosity}. 
$Q_s^\text{visc}$ vanishes for a species undergoing solid-body rotation, when $v_{s\theta} \propto r$. 
Integrated over the cross-sectional area of the system to an outer radius at $r=R$, Eq.~(\ref{eqn:viscousHeating}) becomes 
\begin{align}
2 \pi \int_0^R Q_s^\text{visc} \, r \D r
&= 2 \pi \eta_{s1} r^2 v_{s\theta} \frac{\partial}{\partial r} \bigg( \frac{v_{s \theta}}{r} \bigg) \bigg|_{r=R} \nonumber \\
&\hspace{30 pt}- 2 \pi \int_0^R q_s B v_{s\theta} \Gamma_{s}^\text{visc} \, r \D r. \label{eqn:integratedViscousHeating}
\end{align}
The boundary term in Eq.~(\ref{eqn:integratedViscousHeating}) results from any uncompensated viscous stress at the edges of the system. It vanishes if the rotation profile flattens at the edge, or if a no-slip boundary imposes $v_{s\theta}(R) = 0$. If the dominant rotation is an $\mathbf{E} \times \bB$ drift, then $v_{s \theta} \approx -E/B$, and thus the last term is the Ohmic $\mathbf{j} \cdot \mathbf{E}$ heating due to the current carried by the viscous flux of species $s$. The Ohmic heating term can be rewritten as follows: 
\begin{gather}
- 2 \pi \int_0^R q_s B v_{s \theta} \Gamma_s^\text{visc} r \D r = 2 \pi \int_0^R v_{s \theta} (\nabla \cdot \pi_s)_\theta r \D r. 
\end{gather}
The right-hand side integrand is the dot product of the viscous force with $\bv_s$. 
In other words, the viscous Ohmic dissipation term is exactly compensated by the energy transferred out of the rotational motion by the viscous force. 

In some ways, the two terms on the right-hand side of Eq.~(\ref{eqn:integratedViscousHeating}) -- the boundary term and the Ohmic dissipation -- are independent quantities which can be manipulated separately. For instance, as was noted above, the edge heating can always be eliminated with an appropriate choice of boundary conditions. 
However, these terms are not fully independent. 
Note that the second term on the right-hand side of Eq.~(\ref{eqn:integratedViscousHeating}) can be positive or negative, but that Eq.~(\ref{eqn:viscousHeating}) requires that $Q_s^\text{visc} \geq 0$. This implies that there are cases in which the boundary term in Eq.~(\ref{eqn:integratedViscousHeating}) must be important. 
Any time the viscous heating carries particles from regions of lower electrostatic potential energy to regions of higher potential energy, the boundary term must be larger than the Ohmic heating term. 
For similar reasons, any choice of boundary conditions that eliminates the boundary heating term will implicitly set the sign of $q_s E \Gamma_s^\text{visc}$. 
This point will be significant in Section~\ref{sec:globalConditions}. 

\subsection{Frictional Heating}

The dissipation due to the frictional particle flux $\Gamma_{ss'}^\text{fric}$ behaves quite differently. The heating of species $s$ due to frictional interactions with species $s'$ can be written as 
\begin{align}
Q_{ss'}^\text{fric} &= \frac{m_{s'}}{m_s + m_{s'}} (\bv_{s'}-\bv_s) \cdot \mathbf{R}_{ss'} \label{eqn:generalFrictionalHeating} \\
&= q_s B \bigg( \frac{m_{s'}}{m_s + m_{s'}} \bigg) (v_{s'\theta} - v_{s \theta}) \Gamma_{ss'}^\text{fric} . \label{eqn:frictionalHeating}
\end{align}
This expression satisfies $m_s Q_{ss'}^\text{fric} = m_{s'} Q_{s's}^\text{fric}$. The total frictional heating $Q_s^\text{fric}$ for species $s$ is then 
\begin{gather}
Q_s^\text{fric} = \sum_{s'} Q_{ss'}^\text{fric}. 
\end{gather}

The frictional heating differs from the viscous heating in two key ways. First, though both scale with the size of the corresponding particle flux, the frictional heating depends on the difference in velocities $(v_{s'\theta}-v_{s\theta})$ rather than $v_{s\theta}$ on its own. In an $\bE \times \bB$-rotating device, the rotational velocities for all species will be close to the $\bE\times\bB$ velocity, so $(v_{s' \theta} - v_{s \theta}) \ll v_{s \theta}$. 

Intuitively, why can a viscous particle flux be responsible for so much more dramatic heating than a frictional particle flux of the same size? It follows from the ambipolarity of the frictional flux. The frictional flux $\Gamma_{ss'}^\text{fric}$ of species $s$ interacting with species $s'$ is always paired with a flux $\Gamma_{s's}^\text{fric}$ of species $s'$ interacting with $s$. To leading order, this pair of particle fluxes carries no net charge, and therefore exchanges no net energy with the electrical potential. Exchange of energy with the electrical potential has to come from the higher-order corrections to the flux in Eq.~(\ref{eqn:fluxWithCorrection}). These corrections are discussed in much greater detail elsewhere \cite{Rax2019, Kolmes2019}; they can result in flows of net charge. 

The second key difference between viscous and frictional heating is the way in which heat is divided between the different species. For a viscous particle flux $\Gamma_{s}^\text{visc}$, the associated heating goes entirely into species $s$, although the viscosity coefficient $\eta_{s1}$ can depend on other species. From a macroscopic standpoint, if viscosity drives a current up or down an electrostatic potential, the particles carrying the current pick up (or lose) that potential energy. 
On the other hand, when a frictional interaction between species $s$ and $s'$ drives particle fluxes $\Gamma_{ss'}^\text{fric}$ and $\Gamma_{s's}^\text{fric}$, the associated heating is divided between species $s$ and $s'$ such that each species receives heat inversely proportional to its mass. This happens regardless of which species is actually carrying current. The energy source for heating might be the motion of particles down a potential gradient, but that motion is mediated by collisions between particles, and in those collisions energy is transferred in such a way as to heat the lighter species in any given pair. 

It can be useful to compare the frictional dissipation from cross-field drifts with the Ohmic heating from currents parallel to $\bB$. 
Parallel Ohmic heating is driven by frictional dissipation, so it is reasonably well-described by Eq.~(\ref{eqn:generalFrictionalHeating}) (although a more complete kinetic treatment does lead to corrections \cite{Spitzer1953, Spitzer}). 
In steady state and assuming homogeneity in the parallel direction, the force balance can be written in terms of a parallel field $E_{||}$ as 
\begin{gather}
q_s E_{||} = m_s \sum_{s'} \nu_{ss'} (v_{s||}-v_{s'||}). \label{eqn:parallelForceBalance}
\end{gather}
The velocity difference can be expressed as a function of the species' charge densities, collision frequencies, masses, and $E_{||}$. 
This is quite unlike the perpendicular case, where the $\bE \times \bB$ flows are the same for all species and the velocity differences are instead driven by diamagnetic and inertial effects, and where it is possible to have an electric field without any particle flux or dissipation. 

Combining Eq.~(\ref{eqn:generalFrictionalHeating}) with Eq.~(\ref{eqn:parallelForceBalance}), it follows that so long as the smallest dimensionless parameter in the problem is the electron-to-ion mass ratio, most of the Ohmic heating must go into the electrons. 
This can be understood by considering a system containing two positive ion species, labeled $a$ and $b$. 
So long as $\nu_{ab} \gg \nu_{ae}$, the difference in velocity between species $a$ and $b$ will be much smaller than $|v_{a||}-v_{e||}|$. 
The heating due to interactions between two species scales linearly with the collision frequency but quadratically with the velocity difference, so a pair of species with a high collision frequency will have similar velocities and therefore little heating. 
This is a key difference between parallel and perpendicular frictional dissipation; recall that in the perpendicular case, there are configurations with multiple ion species for which the heating can be directed into the ions. 

\subsection{Frictional Cooling}

There is an odd possibility worth pointing out in Eq.~(\ref{eqn:frictionalHeating}): there are cases in which the frictional heating $Q_{ss'}^\text{fric}$ can be negative. If the frictional force $\mathbf{R}_{ss'}$ acts to reduce the velocity difference between species $s$ and $s'$, it is clear from Eq.~(\ref{eqn:generalFrictionalHeating}) that this cannot happen, and $Q_{ss'}^\text{fric} \geq 0$. But the frictional force can be split into two pieces: a flow friction, which always does act to reduce velocity differences, and a thermal friction, which depends on the temperature gradients and has no particular obligation to align itself with the relative flows of the different species. These two parts of the frictional force can be seen explicitly in Eq.~(\ref{eqn:explicitFriction}). 

At first glance, frictional cooling may be a worrisome thing to find in a transport theory. 
It would be reasonable to wonder if this effect is unphysical. 
On its own, Eq.~(\ref{eqn:generalFrictionalHeating}) contains no obvious fix; $Q_{ss'}^\text{fric}$ and $Q_{s's}^\text{fric}$ always have the same sign, so this is not simply a transfer of energy between species $s$ and $s'$. 
In fact, in order to demonstrate that this frictional cooling is consistent with the second law of thermodynamics, it is necessary to take into account the entropy production from the heat flow $\mathbf{q}_s$. 
The entropy production rate for species $s$ can be written \cite{Hazeltine} as 
\begin{align}
\Theta_s &= \frac{Q_s^\text{fric}}{T_s} + \frac{Q_s^\text{visc}}{T_s} - \frac{\mathbf{q}_s}{T_s} \cdot \frac{\nabla T_s}{T_s} \nonumber \\
&\hspace{50 pt} + \sum_{s'} \frac{3 m_s n_s \nu_{ss'}}{m_s + m_{s'}} \big(T_{s'} - T_s \big). 
\end{align}
The last term is the entropy produced by local temperature equilibration between different species. 
For simplicity, it is helpful to consider the Braginskii single-ion-species limit, in which the frictional heating (and its possible destruction of entropy) affects only the electrons. In this case, 
\begin{align}
\frac{Q_e^\text{fric}}{T_e} 
&= \frac{m_e n_e \nu_{ei}}{T_e} \big(v_{i \theta} - v_{e \theta} \big)^2 \nonumber \\
&\hspace{10 pt}- \frac{3}{2} \frac{m_e n_e \nu_{ei}}{e B T_e} \frac{\partial T_e}{\partial r} (v_{i \theta} - v_{e \theta})
\end{align}
and 
\begin{align}
- \frac{\mathbf{q}_e}{T_e} \cdot \frac{\nabla T_e}{T_e} &= - \frac{3}{2} \frac{m_e n_e \nu_{ei}}{e B T_e} \frac{\partial T_e}{\partial r} (v_{i \theta} - v_{e \theta}) \nonumber \\
&\hspace{10 pt} + 4.66 \, \frac{m_e n_e \nu_{ei}}{e^2 B^2 T_e} \bigg( \frac{\partial T_e}{\partial r} \bigg)^2 . 
\end{align}
The part of the heat flux that depends on the flow velocities is sometimes called the Ettingshausen effect; it is responsible for entropy production or destruction equal to that of the thermal friction. $Q_e^\text{fric} / T_e$ and $-(\mathbf{q}_e \cdot \nabla T_e) / T_e^2$ can each be positive or negative. However, they can be combined as follows: 
\begin{align}
&\frac{Q_e^\text{fric}}{T_e} - \frac{\mathbf{q}_e}{T_e} \cdot \frac{\nabla T_e}{T_e} = 
2.41 \, \frac{m_e n_e \nu_{ei}}{e^2 B^2 T_e} \bigg( \frac{\partial T_e}{\partial r} \bigg)^2 \nonumber \\
&\hspace{50 pt}+\frac{m_e n_e \nu_{ei}}{T_e} \bigg( v_{i \theta} - v_{e \theta} - \frac{3}{2 e B} \frac{\partial T_e}{\partial r} \bigg)^2 . \label{eqn:combinedEntropyProduction}
\end{align}
There are scenarios in which $Q_e^\text{fric} < 0$; these scenarios involve thermal forces driven by temperature gradients. However, for any case in which $Q_e^\text{fric} < 0$, the entropy destroyed by this cooling effect is balanced by entropy produced as heat flows down the temperature gradients. Note that positive entropy production from these mechanisms is not the same as an increase in temperature; the entropy production from the heat flow depends on $\mathbf{q}_s \cdot \nabla T_s / T_s$, whereas the temperature evolution depends on $\nabla \cdot \mathbf{q}_s$. Incidentally, Eq.~(\ref{eqn:combinedEntropyProduction}) provides an alternate way of understanding the numerical instability described in Ref.~\cite{Kolmes2021MITNS}. 
That instability occurred in simulations in which the heat flux was artificially reduced below a certain threshold. Eq.~(\ref{eqn:combinedEntropyProduction}) shows that if the heat flux is artificially suppressed, the entropy production can become negative. 

For the purposes of understanding classical mechanisms for driving temperature differences between species, the possibility of frictional cooling leads to a caveat: $|Q_s^\text{fric}| \gg |Q_{s'}^\text{fric}|$ would not necessarily mean that friction is preferentially heating species $s$. In principle there are temperature and velocity profiles for which friction cools rather than heats. In order for this to happen, the temperature gradients must be large enough for the thermal friction to be larger than the flow friction. 

The frictional cooling is also worth understanding because it explains an otherwise surprising fact that will play a role in Section~\ref{sec:globalConditions}: namely, that $\sum_s Q_s^\text{fric} = 0$ does not necessarily imply that $Q_s^\text{fric} = 0$ for each $s$. 

\section{Controlling the Heating Channels} \label{sec:control}

If the different heating terms associated with cross-field particle transport can be directed into either the ions or the electrons, then under what conditions will ion or electron heating dominate? 
Cross-field viscous dissipation heats ions. 
For a plasma containing a single ion species, the frictional dissipation heats the electrons, so -- apart from the possibility of frictional cooling -- the tendency of classical heating to heat ions or electrons depends entirely on the relative sizes of $Q_i^\text{visc}$ and $Q_e^\text{fric}$. 
For a plasma containing a mix of ion species, ion-ion collisions provide an additional channel for ion heating, and $Q_e^\text{fric}$ must instead compete with $Q_i^\text{visc} + Q_i^\text{fric}$. 

There is more than one way to understand the conditions under which different heating channels will dominate. One approach is to directly compare the expressions for heating given in Eqs.~(\ref{eqn:viscousHeating}) and (\ref{eqn:generalFrictionalHeating}), and to infer the spatial density, velocity, and temperature profiles that maximize or minimize each. 

Consider the conditions under which each of the heating effects vanishes. The viscous dissipation is suppressed when $v_{s \theta} \propto r$ --- that is, in the limit of solid-body rotation. Any time $Q_s^\text{visc}$ vanishes, so does $\Gamma_s^\text{visc}$. However, the reverse is not true; there are profiles with viscous heating but no viscous particle transport. In particular, there is heating without particle transport when 
\begin{gather}
\eta_{s1} r^3 \frac{\partial}{\partial r} \bigg( \frac{v_{s \theta}}{r} \bigg) = \text{nonzero constant}. 
\end{gather}
If $\eta_{s1}$ is spatially constant, this condition corresponds to $v_{s\theta} \propto 1/r$. 

The relationship is the other way around for the frictional heating and transport: there is never nonzero $Q_{ss'}^\text{fric}$ without nonzero $\Gamma_{ss'}^\text{fric}$, but there are profiles with particle transport and no heating. These profiles are discussed in more detail in Appendix~\ref{appendix:profiles}. The particle transport vanishes when $R_{ss'\theta} = 0$. This can be written as a condition on the pressure profiles, because the friction depends on $v_{s \theta}$ and $v_{s' \theta}$, which depend on the species' diamagnetic drifts. If $T_{s'} = \tau T_s$ for some constant $\tau$, $R_{ss'\theta}$ will vanish when 
\begin{align}
&\bigg\{ \frac{p_s(r)}{p_{s}(0)} \exp \bigg[- \int_0^r \D r \bigg( \frac{m_s v_{s \theta}^2}{r T_s} + \gamma_{ss'} T_s'
\bigg) \bigg] \bigg\}^{1/Z_s} \nonumber \\
&\hspace{10 pt}=\bigg\{ \frac{p_{s'}(r)}{p_{s'}(0)} \exp \bigg[ - \int_0^r \D r \bigg( \frac{m_{s'} v_{s' \theta}^2}{r T_{s'}} + \gamma_{s's} T_{s'}' \bigg) \bigg] \bigg\}^{\tau / Z_{s'}} .
\label{eqn:frictionVanish}
\end{align}
In the model used for Eq.~(\ref{eqn:explicitFriction}), 
\begin{gather}
\gamma_{ss'} = \frac{3}{2} \frac{m_{s'}}{m_s T_{s'} + m_{s'} T_s} \, .
\end{gather}
Expressions that are closely related to Eq.~(\ref{eqn:frictionVanish}) have been studied in the particle-transport literature \cite{Spitzer1952, Taylor1961ii, Krishnan1983, Dolgolenko2017, Kolmes2018, Kolmes2020MaxEntropy}. Eq.~(\ref{eqn:frictionVanish}) is more general than the expressions that have been used in the past, since it includes both the effects of the centrifugal potential and the thermal friction, but the generalization is straightforward; see Appendix~\ref{appendix:profiles}. 

The existence of a special class of profiles for which $R_{ss'\theta}$ vanishes is known in the particle transport literature. 
However, it has not been recognized that there is a second special class of dissipationless profiles for which there is particle transport but no frictional heating. 
Note from Eq.~(\ref{eqn:generalFrictionalHeating}) that $Q_{ss'}^\text{fric}$ vanishes whenever either $R_{ss'\theta} = 0$ or $v_{s \theta} = v_{s'\theta}$. If $T_{s'} = \tau T_s$, then $v_{s\theta} = v_{s'\theta}$ when 
\begin{align}
&\bigg\{ \frac{p_s(r)}{p_{s}(0)} \exp \bigg[- \int_0^r \D r \, \frac{m_s v_{s \theta}^2}{r T_s} \bigg] \bigg\}^{1/Z_s} \nonumber \\
&\hspace{30 pt}=\bigg\{ \frac{p_{s'}(r)}{p_{s'}(0)} \exp \bigg[ - \int_0^r \D r \, \frac{m_{s'} v_{s' \theta}^2}{r T_{s'}} \bigg] \bigg\}^{\tau / Z_{s'}} .
\label{eqn:deltaVVanish}
\end{align}
This condition is derived in Appendix~\ref{appendix:profiles}. 
The profiles described by Eq.~(\ref{eqn:frictionVanish}) and those described by Eq.~(\ref{eqn:deltaVVanish}) become the same when the temperature gradients vanish, or more generally when 
\begin{gather}
\frac{m_{s'} T_s T_s'}{q_s} = \frac{m_s T_{s'} T_{s'}'}{q_{s'}} \, .
\end{gather}

The frictional heating $Q_{ss'}^\text{fric}$ can be understood as a function of how close the different pressure profiles are to the classes of dissipationless profiles described by Eqs.~(\ref{eqn:frictionVanish}) and (\ref{eqn:deltaVVanish}). A pair of species that satisfies Eq.~(\ref{eqn:deltaVVanish}) will have a cross-field frictional particle flux but no corresponding heating. A pair that satisfies Eq.~(\ref{eqn:frictionVanish}) will have no cross-field frictional particle flux (and no corresponding heating). 

Qualitatively, the relative importance of viscous and frictional heating depends on how far the velocity profile is from solid-body rotation compared with how far the pressure profiles are from satisfying Eqs.~(\ref{eqn:frictionVanish}) or (\ref{eqn:deltaVVanish}). 
Similarly, the relative importance of frictional ion-ion and ion-electron heating can be understood in terms of the deviations of the different species' profiles from Eqs.~(\ref{eqn:frictionVanish}) and (\ref{eqn:deltaVVanish}), weighted by the appropriate collision frequencies. For species with comparable densities, ion-ion collision frequencies are larger than ion-electron collision frequencies by a factor of the square root of the ion-electron mass ratio. As such, in order for electron heating to dominate, not only must the velocity profile not be too far from solid-body rotation, but the electrons must be substantially further away from satisfying Eqs.~(\ref{eqn:frictionVanish}) or (\ref{eqn:deltaVVanish}) with respect to the ions than the ions are with respect to other ion species. 

\section{Global Conditions for Ion Heating} \label{sec:globalConditions}

Remarkably, there are circumstances under which it is possible to determine which heating terms dominate without knowing anything other than the global 0-D particle balance and the boundary conditions. 
This is to be contrasted with the approach taken in the previous section, which relied on knowledge of the full spatial profiles of density, velocity, and temperature. 

Consider a simple model of a cylindrical fusion device with total radius $R$. 
Pick boundary conditions at the radial boundary $r=R$, such that either $v_{s\theta}$ or $\partial (v_{s \theta} / r) / \partial r$ vanishes at the boundary (so that the boundary term in Eq.~(\ref{eqn:integratedViscousHeating}) can be ignored). 
From Eq.~(\ref{eqn:integratedViscousHeating}) and the fact that $Q_s^\text{visc} \geq 0$, either boundary condition implies that $q_s \Gamma_s^\text{visc} E_r \geq 0$. 

\begin{figure}[h]
	\centering
	\begin{tikzpicture} 
	
	\def\r{1.};
	\def\R{8};
	\def\a{20.};
	
	\draw[thick, black, dashed] ({\r*cos(-\a)},{\r*sin(-\a)}) arc (-\a:\a:\r);
	\draw[thick, black] ({\R*cos(-\a)},{\R*sin(-\a)}) arc (-\a:\a:\R);
	
	\draw[thick, black, dashed] ({\r*cos(-\a)},{\r*sin(-\a)}) -- ({\R*cos(-\a)},{\R*sin(-\a)}); 
	\draw[thick, black, dashed] ({\r*cos(\a)},{\r*sin(\a)}) -- ({\R*cos(\a)},{\R*sin(\a)});
	
	\draw[ultra thick, orange, decorate, decoration=snake, ->] ({(\r+.15)*cos(-.5*\a)},{(\r+.15)*sin(-.5*\a)}) -- ({(\R-.15)*cos(-.5*\a)},{(\R-.15)*sin(-.5*\a)});
	
	\draw[ultra thick, blue, <-] ({(\r+.15)*cos(.5*\a)},{(\r+.15)*sin(.5*\a)}) -- ({(\R-.15)*cos(.5*\a)},{(\R-.15)*sin(.5*\a)});
	
	\node at (7,1.8) {fuel ions};
	\node at (6.6,-1.8) {fusion products};
	
	\draw[fill, yellow] (0,0) circle (.55);
	\node at (0,0) {fusion};
	
	
	\end{tikzpicture}
	\caption{The global particle balance for a cylindrical system with fusion reactions taking place at the core. If an effect like $\alpha$-channeling removes the fusion products, classical transport must provide a balancing influx of fuel ions. } \label{fig:particleBalance} 
\end{figure}
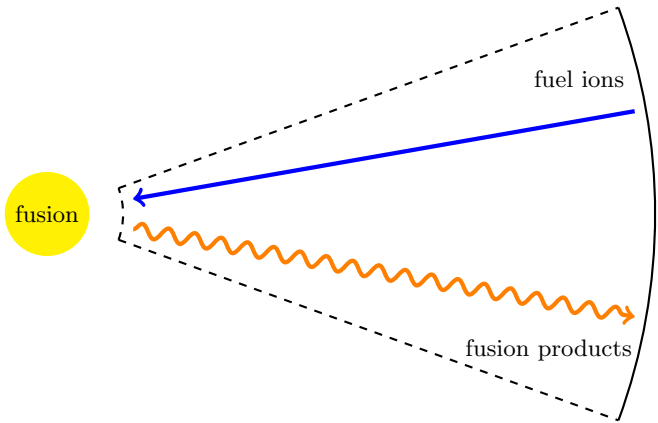

Suppose fuel ions $a$ and $b$ are supplied at the edge and suppose there is a fusion reaction consuming one ion of species $a$ and one ion of species $b$ occurring at some rate $S$ per unit length at $r = 0$. 
Finally, suppose the fusion products are removed promptly, by a process other than classical transport, which we can imagine to be wave-driven, as in $\alpha$-channeling \cite{Fisch1992}. 
Suppose the fusion products are removed before they interact with the rest of the plasma, collisionally or otherwise \cite{Chen2016}. 
Then, in order to achieve steady state, there must be injection of fuel ions into the system. 
This is shown schematically in Figure~\ref{fig:particleBalance}. 

First assume that the wave-driven removal of the fusion ash is non-ambipolar, in the sense that only the ash ions (and not an accompanying population of electrons) are removed. 
This is the usual circumstance envisioned in the case of $\alpha$-channeling, where resonant wave-particle interactions promptly  eject the $\alpha$ particles on a collisionless timescale. 
Then the total fuel ion fluxes must satisfy 
\begin{gather}
\Gamma_a = \Gamma_b = - \frac{S}{2 \pi r} \label{eqn:nonambipolarIonFluxCondition}
\end{gather}
for all $r \in (0,R)$, and the electron flux is
\begin{gather}
\Gamma_e = 0. \label{eqn:zeroElectronFluxCondition}
\end{gather}
When new fuel ions are supplied at the edge, they must be at rest in order to ensure that the total angular momentum in the system remains constant; the $\mathbf{j}\times\bB$ torques due to the inward motion of the fuel ions and the outward motion of the ash will balance one another. 

The fluxes given by Eqs.~(\ref{eqn:nonambipolarIonFluxCondition}) and (\ref{eqn:zeroElectronFluxCondition}) will in general be a mix of frictional and viscous transport. 
However, almost all of the net flow of charge must be carried by the viscous fluxes $\Gamma_a^\text{visc}$ and $\Gamma_b^\text{visc}$. 
This is not at all obvious. 
Note that we have not made any assumptions about the relative sizes of $\Gamma_s^\text{visc}$ and $\Gamma_{ss'}^\text{fric}$ for the ions, so one might imagine a case in which $\Gamma_s^\text{visc}$ is at least $\mathcal{O}(\delta)$ smaller than $\sum_{s'} \Gamma_{ss'}^\text{fric}$ and the current is instead carried by the $\mathcal{O}(\delta)$ terms in Eq.~(\ref{eqn:fluxWithCorrection}). 
Indeed, if the $\mathcal{O}(\delta)$ corrections were entirely arbitrary, this might be possible. But they are not arbitrary; recall that Eq.~(\ref{eqn:radialFlux}) can be expanded as 
\begin{gather}
\Gamma_s = \bigg( \Gamma_s^\text{visc} + \sum_{s'} \Gamma_{ss'}^\text{fric} \bigg) \sum_{k=0}^\infty \bigg[ - \frac{(r v_{s \theta})'}{r \Omega_s} \bigg]^k . \label{eqn:radialFluxWithSum}
\end{gather}
Intuitively, the $(k+1)$th term in the sum in Eq.~(\ref{eqn:radialFluxWithSum}) can be understood as the $\mathbf{F} \times \mathbf{B}$ flow from the Coriolis and Euler forces that result from the radial motion described by the $k$th term \cite{Kolmes2019}.

The $k \neq 0$ corrections in Eq.~(\ref{eqn:radialFluxWithSum}) can produce non-ambipolar corrections to otherwise ambipolar flows. 
However, a non-ambipolar correction will only appear if there is a non-vanishing total flow at the next-lowest order. 
In other words, the $k \neq 0$ terms can contribute a significant part of the current-carrying flow only if there is a non-vanishing ambipolar flow that is much larger. 
In this scenario, the leading-order flow is already non-ambipolar, so the flow of charge cannot be explained by these corrections. 

As such, to leading order in $\delta$, it follows that the current is carried by the viscous fluxes: 
\begin{gather}
Z_a \Gamma_a^\text{visc} + Z_b \Gamma_b^\text{visc} = - (Z_a + Z_b) \frac{S}{2 \pi r}. \label{eqn:viscousCurrentWins}
\end{gather}
Then, up to $\mathcal{O}(\delta)$ corrections, the heating associated with the flow of charge down the electrical potential goes into the ion species. 

Even so, there could be large frictional fluxes in the system; if there are temperature gradients, then it is possible to have $\Gamma_s^\text{fric} = 0$ but $\Gamma_{ss'}^\text{fric} \neq 0$. This can happen when the thermal friction is at least comparable to the friction from flows. In that case, it is possible to have $Q_{ss'}^\text{fric} \neq 0$. If $\Gamma_s^\text{fric} = 0$ for all species, then 
\begin{align}
&\sum_{s,s'} Q_{ss'}^\text{fric} = \sum_{s,s'} q_s B \bigg( \frac{m_{s'}}{m_s + m_{s'}} \bigg) (v_{s' \theta} - v_{s \theta}) \Gamma_{ss'}^\text{fric} \\
&= - \frac{B}{2} \bigg\{ \sum_s q_s v_{s \theta} \sum_{s'} \Gamma_{ss'}^\text{fric} + \sum_{s'} q_{s'} v_{s' \theta} \sum_s \Gamma_{s's}^\text{fric} \bigg\} \\
&= 0. 
\end{align}
The frictional heating $\sum_{s'} Q_{ss'}^\text{fric}$ for any given species $s$ does not necessarily vanish on its own, so it must represent a transfer of heat between species (recall that $Q_{ss'}^\text{fric}$ can be frictional heating or frictional cooling). However, if the thermal friction is on the same order as the total friction, it is possible to show that $\sum_{s'} Q_{ss'}^\text{fric}$ is very small compared to the temperature equilibration term between any two species, so the energy transfer associated with the frictional flows will not have a significant effect on the relative temperatures of the different species. 

Eq.~(\ref{eqn:viscousCurrentWins}) is an interesting result because it implies that, for every fusion event, the overall population of fuel ions would be heated by $(Z_a + Z_b) e \Delta \Phi$, where $\Delta \Phi$ is the potential difference between the edge and the center. 
This is possible because, for these boundary conditions, all of the potential energy liberated by a viscosity-driven current (i.e., the full $\mathbf{j} \cdot \mathbf{E}$) goes into the ions. 
For a system with very large fields, this would represent a very large ion heating effect. 

Note that the scenario put forth in this section is a special case that relies on the transport being non-ambipolar to leading order in the ratio of the rotation frequency to the ion gyrofrequency.  
This ordering can be thought of as a limit to the electric field (since $\delta \propto E$), which in turn limits only how much energy can be gained by falling through the electric potential. 
Note also that in a case in which the net classical transport were ambipolar (for instance, if the mechanism that removed the fusion ash also removed a balancing population of electrons), it would not be possible to determine the dominant heating channel through global particle balance constraints alone. 
Instead, it would be necessary to consider the full spatial profiles, along the lines of the discussion in Section~\ref{sec:control}. 

\section{Example Configurations} \label{sec:examples}

There are a couple of different proposed fusion reactor configurations with particularly large voltage drops between the edge and the core of the plasma, such that the ion $\mathbf{j} \cdot \mathbf{E}$ dissipation described in Section~\ref{sec:control} could be very significant. For instance, the Wave-Driven Rotating Torus (or WDRT) is a proposed toroidal configuration in which a minor-radial electric field and toroidal mangetic field produce sufficiently large $\bE \times \bB$ flows to produce a rotational transform \cite{Rax2017}, essentially fulfilling the same function as the poloidal field in a tokamak. A WDRT would likely involve MV-scale potential differences between the edge of the system and the core. 

Comparable voltage drops could be found in a rotating magnetic mirror: that is, a linear system in which the axial confinement is enhanced by the centrifugal force from fast rotation \cite{Lehnert1971, Bekhtenev1980, Ellis2001, Ellis2005, Teodorescu2010}. In this case, the  expulsion of fusion products via $\alpha$-channeling could be arranged with a stationary magnetic field perturbation \cite{Fetterman2010, Fetterman2011}, removing the need to inject RF waves. For example, Bekhtenev suggested a configuration with a density of about $3 \times 10^{13} \text{ cm}^{-3}$, ion temperatures between 30 and 100 keV, rotational energy around 150 - 500 keV, and a central magnetic field strength between 1.5 and 2.5~T. For a 2~T field, simulations by Fetterman and Fisch suggest that the $\alpha$-channeling effect could be accomplished with a stationary field ripple of 0.1~T \cite{Fetterman2011c}. 

In either configuration, it would be reasonable to expect $\mathcal{O}(1)$~MV voltage drops across the system. 
In such a case, the $\mathbf{j} \cdot \bE$ dissipation could deliver $\mathcal{O}(Z_a + Z_b) \text{ MeV}$ to the fuel ions for every fusion event. The total power going into the ions through classical viscous dissipation could be on the same order as the part of the fusion power that resides in charged particles. The overall power balance of the system would involve some significant fraction of the fusion power being diverted through $\alpha$-channeling into maintaining the electrostatic potential, and fuel ions taking large amounts of energy from that potential as they move from edge to core. 
Of course, the mechanism described in Section~\ref{sec:control} does not require MV-scale potential differences. For smaller $\Delta \Phi$ (that is, slower rotation), all other things being equal, the size of the $\mathbf{j} \cdot \mathbf{E}$ dissipation (and the associated tendency of the ion temperature to exceed the electron temperature) would be reduced. 

\section{Caveats} \label{sec:realizability}

The foregoing analysis relies on a number of assumptions about the behavior of the system. 
It is reasonable to wonder how easy or difficult each of these would be to realize in practice. 

First, the transport model used here assumes that cross-field particle transport is classical. 
This may be the most difficult condition to realize; after all, turbulent transport dominates in many fusion experiments, often by a large margin. 
However, there are strong indications that classical transport may be more easily attainable in rotating plasmas, and that sheared rotation can suppress turbulence \cite{Taylor1989, Cho2005, Maggs2007}. 
A fully classical fusion device would certainly represent a significant technical and scientific achievement, but there are reasons to be optimistic. 
Of course, it may be that there are regimes in which turbulent transport can also produce dissipation in a way that leads to a hot ion mode; indeed, turbulent transport can also produce dissipation, especially when it involves net motion aligned with an electric field \cite{Hinton2006, Garbet2013}. This goes beyond the scope of this paper, but it could be worth future investigation. 

Second, the calculations in this paper assume that axial effects can be neglected. 
In principle, even in the limit of very poor axial confinement, this problem could be circumvented by a very long device \cite{Ivanov2013}. 
Again, though, the presence of rotation makes this condition more realizable. 
In mirror-type systems, rotation can improve axial confinement. 
If the rotation is sufficiently fast, this improvement can be dramatic \cite{Lehnert1971, Bekhtenev1980}, though rotation does not fully eliminate axial losses. 

Third, in the case of the 0-D fusion reactor model presented in Section~\ref{sec:globalConditions}, this paper assumes that there is a mechanism like $\alpha$-channeling that can remove fusion products with incurring collisional effects, and that this mechanism does not also expel electrons or draw in fuel ions. $\alpha$-channeling has a good theoretical basis \cite{Fisch1992,Valeo1994,Fisch1995ii,Fisch1995,Ochs2015b,Cianfrani2018,Cianfrani2019,Castaldo2019,Romanelli2020}, and aspects of the theory have been validated by experiment \cite{Darrow1996, Fisch2000}, but full experimental validation remains an open problem. 

Fourth, the reactor model in Section~\ref{sec:globalConditions} assumes a simple boundary condition on the rotation profile (that either the rotation or the shear vanish on the boundary). A fully realistic treatment of the boundary would need to include things like plasma-surface and plasma-neutral interactions, which are neglected here. 

Finally, Section~\ref{sec:globalConditions} assumes that the plasma is in a steady state with fusion reactions taking place near $r = 0$. The results of the analysis would be different if the system were pulsed, at least if the plasma lifetime was not long compared to the timescales of cross-field transport and fuel depletion. 

\section{Discussion} \label{sec:discussion}

In a rotating plasma, the classical dissipation can proceed through several very different mechanisms. 
By controlling which mechanism dominates, the power flow through the plasma can be altered significantly. 
If the right channels dominate -- and if the dissipation is large -- it is possible to create what we have called a ``natural" hot ion mode, produced without any need for auxiliary heating of the ions. 

In the most general case, the heating going through each of the classical dissipation channels depends in a complicated way on the density, rotation velocity, and temperature of each species. 
However, the viscous heating is large when the velocity profiles are far from solid-body rotation. 
The frictional heating is large when the pressure profiles deviate from the set of pairwise relations to which the system would in principle relax, were it not driven. 
Thus, ions are preferentially heated when viscous dissipation and ion-ion frictional dissipation dominate ion-electron frictional dissipation. 
In contrast, in the case of parallel Ohmic dissipation, the ion-electron friction always dominates, so that it is always the electrons that are heated. 
A related set of concerns has been considered in the neutral-beam heating literature by Helander, Akers, and Eriksson; they pointed out that viscous heating of ions can be significant in that context \cite{Helander2005}. 

It is quite remarkable that, in a simple 0-D model of a fusion reactor, it is possible to determine the dominant heating channel through global constraints alone. 
In particular, with appropriate constraints on the boundary conditions and the behavior of the fusion products, global constraints, that force the particle fluxes to be non-ambipolar to leading order in rotation frequency over ion gyrofrequency, also ensure that heat flows to the ions through viscous dissipation. 
The ions then receive energy in the form of heat comparable to the electrostatic potential energy drop from the edge to the core.
This ion heating may then produce a hot-ion mode significant enough to facilitate economical nuclear fusion. 

There are scenarios in which the waves used for $\alpha$-channeling could damp on the fuel ions directly (for instance, if the amplified waves encounter the tritium resonance \cite{Valeo1994}). 
This would increase $T_i - T_e$ significantly, and could circumvent the need for classical dissipation to produce a hot-ion mode. However, if $\alpha$-channeling does move net charge across field lines, and if $E_r < 0$, then the wave would no longer absorb all of the kinetic energy from the fusion products, because pulling fusion products out of an electrostatic potential comes with an energetic cost \cite{Fetterman2008, Fetterman2010, Fetterman2011}. If the potential is large, much of the energetic effect of $\alpha$-channeling could be simply to transfer the fusion energy to the potential. This would tend to make the classical dissipation mechanisms discussed here more important, since it reduces the amount of available energy in the wave that could heat the fuel ions directly. Taken together, the net effect would be to move energy from the fusion products to the fuel ions, effectively replacing the conventional $\alpha$-particle slowing down process with a mechanism that does not heat electrons. 

Regardless of precisely what happens to the wave energy, what we have shown here is that in a rotating hot plasma, in which fusion ash is expelled promptly while fuel ions are drawn in through collisional transport, there is a tendency for the natural dissipation in maintaining the plasma configuration to favor heat going into the ions rather than the electrons. 
This produces the possibility of a naturally occurring hot-ion mode, without the need for external mechanisms that directly heat fuel ions. 
This remarkable circumstance could be of great interest in economical controlled nuclear fusion.

~\\
\begin{acknowledgements}
The authors would like to thank S. Jin and T. Rubin for helpful conversations. 
This work was supported by
Cornell NNSA 83228-10966 [Prime No.~DOE (NNSA) DENA0003764]
and by NSF-PHY-1805316.
\end{acknowledgements}

\begin{appendix}

\section{Viscous Forces and Heating} \label{appendix:viscosity}

Consider a system with cylindrical symmetry, no flow in the $\hat z$ direction, and a magnetic field in the $\hat z$ direction. The velocity of species $s$ can be written as 
\begin{gather}
\bv_s = v_{sr}(r) \hat r + r \omega_s(r) \hat \theta. \label{eqn:velocityProfile}
\end{gather}
Suppose the plasma is weakly coupled and strongly magnetized; the viscosity can change significantly if either of these conditions is not met \cite{Scheiner2020}. Moreover, suppose Braginskii's flow ordering is satisfied, so that the flow velocities are comparable to the ion thermal velocity. 

For the following calculation, the species index $s$ will be suppressed. The viscosity tensor $\pi_{ij}$ can be calculated in an arbitrary coordinate system using the procedure outlined in Ref.~\cite{Kolmes2019}. It is convenient to pick cylindrical coordinates, for which the metric tensor $g_{ij}$ is 
\begin{gather}
g_{ij} = \begin{pmatrix} 1 & 0 & 0 \\ 0 & r^2 & 0 \\ 0 & 0 & 1 \end{pmatrix}
\end{gather}
and the Christoffel symbols are given by 
\begin{align}
\Gamma^r_{ij} &= \begin{pmatrix} 0 & 0 & 0 \\ 0 & -r & 0 \\ 0 & 0 & 0 \end{pmatrix} \\
\Gamma^\theta_{ij} &= \begin{pmatrix} 0 & r^{-1} & 0 \\ r^{-1} & 0 & 0 \\ 0 & 0 & 0 \end{pmatrix} \\
\Gamma^z_{ij} &= 0. 
\end{align}
Define a covariant velocity vector $\mathbf{u}$ in this coordinate system, satisfying $u_r = v_r$, $u_\theta = r^2 \omega$, and $u_z = 0$. By definition, the corresponding contravariant vector has components $u^i = g^{ij} u_j$, so $u^r = v_r$, $u^\theta = \omega$, and $u^z = 0$. 

In terms of the covariant derivative $\nabla_i$, Braginskii's traceless rate-of-strain tensor $W_{ij}$ is defined by 
\begin{align}
W_{ij} \doteq \nabla_i u_j + \nabla_j u_i - \frac{2}{3} g_{ij} \nabla_k u^k . 
\end{align}
For the velocity given by Eq.~(\ref{eqn:velocityProfile}), its nonzero components are 
\begin{align}
W_{rr} &= \frac{2}{3} \big( 2 \partial_r u_r - r^{-1} u_r \big) \\
W_{r \theta} &= W_{\theta r} = \partial_r u_\theta - 2 r^{-1} u_\theta \\
W_{\theta \theta} &= - \frac{2}{3} \big( r^2 \partial_r u_r - 2 r u_r \big) \\
W_{zz} &= - \frac{2}{3} \big( \partial_r u_r + r^{-1} u_r \big) . 
\end{align}
Braginskii's viscosity tensor can be written as 
\begin{align}
\pi_{ij} &= - \eta_0 \, ^0 W_{ij} - \eta_1 \, ^1 W_{ij} - \eta_2 \, ^2W_{ij} \nonumber \\
&\hspace{10 pt} + \eta_3 \, ^3 W_{ij} + \eta_4 \, ^4W_{ij} , \label{eqn:piAndEtas}
\end{align}
where the $^k W$ tensors can be written in covariant form \cite{Kolmes2019} as
\begin{align}
&^0 W_{ij} = \frac{3}{2} \big( b_i b_j - \frac{g_{ij}}{3} \big) \big( b_m b_n - \frac{g_{mn}}{3} \big) W^{mn} \\
&^1 W_{ij} = \big( \delta_{im}^\perp \delta_{nj}^\perp + \frac{1}{2} \delta_{ij}^\perp b_m b_n \big) W^{mn} \\
&^2 W_{ij} = \big( \delta_{im}^\perp b_j b_n + \delta_{nj}^\perp b_i b_m \big) W^{mn} \\
&^3 W_{ij} = \frac{1}{2} \big( \delta_{im}^\perp \epsilon_{nbk}b^k - \delta_{nj}^\perp \epsilon_{imk}b^k \big) W^{mn} \\
&^4 W_{ij} = \big( b_i b_m \epsilon_{njk}b^k + b_j b_n \epsilon_{imk}b^k \big) W^{mn} . 
\end{align}
Here $b_i = \begin{pmatrix} 0 & 0 & 1 \end{pmatrix}$ is the unit vector in the direction of the magnetic field and $\delta_{ij}^\perp \doteq g_{ij} - b_i b_j$. In this coordinate system, the Levi-Civita tensor is defined as 
\begin{gather}
\epsilon_{ijk} \doteq r \tilde{\epsilon}_{ijk}, 
\end{gather}
where $\tilde{\epsilon}_{ijk}$ is the Levi-Civita symbol. 

These expressions are derived from Braginskii's model, which assumes a single ion species. For a plasma containing several ion species, the viscosity can still be modeled using Eq.~(\ref{eqn:piAndEtas}), but the expressions for the $\eta$ coefficients for each species are modified \cite{Zhdanov}. 

With these definitions, it is possible to directly calculate $\pi_{ij}$ in cylindrical coordinates for the chosen velocity profile: 
\begin{align}
\pi_{ij} &= - \frac{\eta_0}{3 r} \frac{\partial (r u_r)}{\partial r} \begin{pmatrix} 
1 & 0 & 0 \\
0 & r^2 & 0 \\
0 & 0 & -2 \end{pmatrix} \nonumber \\
&\hspace{5 pt} - \bigg[ \eta_{1} r \frac{\partial}{\partial r} \bigg( \frac{u_r}{r} \bigg) + \eta_3 r \frac{\partial}{\partial r} \bigg( \frac{u_\theta}{r^2} \bigg) \bigg] \begin{pmatrix}
1 & 0 & 0 \\
0 & -r^2 & 0 \\
0 & 0 & 0
\end{pmatrix} \nonumber \\ 
&\hspace{5 pt} - \bigg[ \eta_1 r^2 \frac{\partial}{\partial r} \bigg( \frac{u_\theta}{r^2} \bigg) - \eta_3 r^2 \frac{\partial}{\partial r} \bigg( \frac{u_r}{r} \bigg) \bigg] \begin{pmatrix}
0 & 1 & 0 \\
1 & 0 & 0 \\
0 & 0 & 0
\end{pmatrix} . \label{eqn:matrixGroupings}
\end{align}
Then the viscous force density can be written as 
\begin{align}
\nabla_i \pi^{ij} &= \partial_i \pi^{ij} + \Gamma_{i \lambda}^i \pi^{\lambda j} + \Gamma_{i \lambda}^j \pi^{i \lambda} . 
\end{align}
The first and second of the three matrices in Eq.~(\ref{eqn:matrixGroupings}) contribute to the $\hat r$-directed force. After converting from the resulting contravariant force vector back to the original coordinate normalization, 
\begin{align}
(\nabla \cdot \pi)_r &= - \frac{\partial}{\partial r} \bigg[ \frac{\eta_0}{3 r} \frac{\partial (r v_r)}{\partial r} \bigg] \nonumber \\
&\hspace{20 pt}- \frac{1}{r^2} \frac{\partial}{\partial r} \bigg[ \eta_1 r^3 \frac{\partial}{\partial r} \bigg( \frac{v_r}{r} \bigg) + \eta_3 r^3 \frac{\partial \omega}{\partial r} \bigg] . \label{eqn:rViscousForce}
\end{align}
Only the third matrix in Eq.~(\ref{eqn:matrixGroupings}) contributes to the $\hat \theta$-directed force. Again, in the original normalization, 
\begin{gather}
(\nabla \cdot \pi)_\theta = - \frac{1}{r^2} \frac{\partial}{\partial r} \bigg[ \eta_1 r^3 \frac{\partial \omega}{\partial r} - \eta_3 r^3 \frac{\partial}{\partial r} \bigg( \frac{v_r}{r} \bigg) \bigg] . \label{eqn:thetaViscousForce}
\end{gather}
For the chosen velocity profile, there is no viscous force in the $\hat z$ direction. 

The relative importance of these different terms depends, in part, on the $\eta$ coefficients. From this point forward, the species indices will no longer be suppressed. For a plasma containing a single ion species, Braginskii gives ion coefficients \cite{Braginskii1965}
\begin{align}
\eta_{i0} &= 0.96 \sqrt{2} \, \frac{n_i T_i}{\nu_{ii}} \\
\eta_{i1} &= \frac{3}{10 \sqrt{2}} \frac{n_i T_i \nu_{ii}}{\Omega_i^2} \\
\eta_{i3} &= \frac{1}{2} \frac{n_i T_i}{\Omega_i}
\end{align}
and electron coefficients
\begin{align}
\eta_{e0} &= 0.73 \, \frac{n_e T_e}{\nu_{ei}} \\
\eta_{e1} &= 0.51 \, \frac{n_e T_e \nu_{ei}}{\Omega_e^2} \\
\eta_{e3} &= \frac{1}{2} \frac{n_e T_e}{|\Omega_e|} \, ,
\end{align}
where $\Omega_s \doteq q_s B / m_s$. In the multiple-ion-species case, the coefficients scale similarly \cite{Zhdanov}, but in general there are additional contributions due to collisions with other species. 

If the radial flow is driven by classical transport, then $v_{sr}$ will be much smaller than $v_{s \theta}$. 
In the case of a plasma with a single ion species \cite{Kolmes2019},
\begin{align}
\frac{v_{ir}}{v_{i \theta}} \sim \frac{v_{er}}{v_{e \theta}} \sim \frac{\nu_{ie}}{\Omega_i} \frac{E}{r B \Omega_i} \, . \label{eqn:velocityComponentScaling}
\end{align}
This is because the frictional $\mathbf{F} \times \bB$ flow due to the difference between ion and electron azimuthal velocities is smaller than the azimuthal velocity difference by a factor of $\nu_{ie} / \Omega_i$, and the azimuthal velocity difference is typically small compared to the total azimuthal velocity by a factor of $E / r B \Omega_i$. In the multiple-ion-species case, the radial transport could instead be driven by unlike-ion collisions, in which case the radial flow would be larger by a factor of $\mathcal{O}(\nu_{ii'}/\nu_{ie})$ than what is given in Eq.~(\ref{eqn:velocityComponentScaling}). Meanwhile, $\eta_{i1} / \eta_{i3} \sim \nu_{ii} / \Omega_i$. For this reason, for the flows studied in this paper, Eq.~(\ref{eqn:thetaViscousForce}) can be approximated as 
\begin{gather}
(\nabla \cdot \pi_i)_\theta \approx - \frac{1}{r^2} \frac{\partial}{\partial r} \bigg[ \eta_{i1} r^3 \frac{\partial}{\partial r} \bigg( \frac{v_{i \theta}}{r} \bigg) \bigg] .
\end{gather}
The relative sizes of the $\eta$ coefficients also imply that $(\nabla \cdot \pi_e)_\theta$ is small compared to $(\nabla \cdot \pi_i)_\theta$. 

The radial viscous forces may be relatively large. The largest of the $\eta$ coefficients to appear in Eq.~(\ref{eqn:rViscousForce}) is $\eta_{s0}$. However, note that the part of $(\nabla \cdot \pi_s)_r$ that depends on $\eta_{s0}$ will vanish for any divergenceless radial flow. Moreover, if the flow is not divergenceless, the part of the radial force that depends on $\eta_{s0}$ scales like 
\begin{align}
- \frac{\partial}{\partial r} \bigg[ \frac{\eta_{i0}}{3r} \frac{\partial (r v_{ir})}{\partial r} \bigg] &\sim \frac{v_{ir}}{\nu_{ii} L} \frac{p_i}{L} \\
- \frac{\partial}{\partial r} \bigg[ \frac{\eta_{e0}}{3r} \frac{\partial (r v_{er})}{\partial r} \bigg] &\sim \frac{v_{er}}{\nu_{ie} L} \frac{p_e}{L} \, ,
\end{align}
where $L$ is the characteristic gradient scale length. Radial flows driven by classically transport will typically move particles much less than the gradient scale length over the course of a collision time. Therefore, this part of the radial viscous force is negligible compared to the pressure force. Similar arguments apply for the other terms in Eq.~(\ref{eqn:rViscousForce}); after all, $\eta_{s1}$ and $\eta_{s3}$ are small compared to $\eta_{s0}$. 

Note that this scaling argument should not be used to neglect the $\hat \theta$ component of the viscous force, since it is in a different direction. In an axially magnetized plasma, azimuthal forces are much more efficient than radial forces at driving radial transport, because azimuthal forces produce $\mathbf{F} \times \mathbf{B}$ drifts in the radial direction. This is discussed in greater detail in Ref.~\cite{Kolmes2019}. 

The other important effect of viscosity is the viscous heat dissipation. The viscous heating for species $s$ is 
\begin{align}
Q_s^\text{visc} &= - \pi_s^{ij} \nabla_i u_j \, ,
\end{align}
which can be evaluated as 
\begin{align}
Q_s^\text{visc} &= \frac{\eta_{s0}}{3 r^2} \bigg[ \frac{\partial (r v_{sr})}{\partial r} \bigg]^2 + \eta_{s1} r^2 \bigg[ \frac{\partial}{\partial r} \bigg( \frac{v_{sr}}{r} \bigg) \bigg]^2 \nonumber \\
&\hspace{85 pt} + \eta_{s1} r^2 \bigg[ \frac{\partial}{\partial r} \bigg(\frac{v_{s \theta}}{r} \bigg) \bigg]^2 . \label{eqn:fullViscousHeating}
\end{align}
The ion $\eta$ coefficients are large compared to the corresponding electron coefficients, so $Q_i^\text{visc} \gg Q_e^\text{visc}$. 
For a divergenceless flow, the part of $Q_i^\text{visc}$ that depends on $\eta_{i0}$ vanishes. If the flow is not divergenceless, then 
\begin{align}
&\frac{\eta_{i0}}{3 r^2} \bigg[ \frac{\partial (r v_{ir})}{\partial r} \bigg]^2 \bigg/ \eta_{i1} r^2 \bigg[ \frac{\partial}{\partial r} \bigg( \frac{v_{i \theta}}{r} \bigg) \bigg]^2 \nonumber \\
&\hspace{140 pt}\sim \bigg( \frac{\Omega_i}{\nu_{ii}} \frac{v_{ir}}{v_{i \theta}} \bigg)^2 . \label{eqn:viscRRatio}
\end{align}
According to the scaling in Eq.~(\ref{eqn:velocityComponentScaling}), this ratio is small, and the first of the three terms in Eq.~(\ref{eqn:fullViscousHeating}) can be neglected. The second of the three terms can be neglected whenever the radial velocity is small compared to the azimuthal velocity, as will be the case if that radial flow is driven by classical transport. Therefore, the viscous heating for ion species $i$ can reasonably be approximated by 
\begin{gather}
Q_i^\text{visc} \approx \eta_{i1} r^2 \bigg[ \frac{\partial}{\partial r} \bigg( \frac{v_{i \theta}}{r} \bigg) \bigg]^2. 
\end{gather}
This is the expression used elsewhere in the paper. Because of the symmetric geometry and velocity profile used here, this treatment includes only the perpendicular viscosity; however, note that there are contexts (particularly involving MHD fluctuations) in which the parallel viscosity may drive significant ion heating \cite{Gimblett1990, Sasaki1997, Haines2006}. 

\section{Derivation of Dissipationless Pressure Profiles} \label{appendix:profiles}

The frictional heating of species $s$ due to frictional interactions with species $s'$ can be written as 
\begin{gather}
Q_{ss'}^\text{fric} = \frac{m_{s'}}{m_s + m_{s'}} \big( \bv_{s'} - \bv_s \big) \cdot \mathbf{R}_{ss'}. 
\end{gather}
In the typical case where $v_{sr} \ll v_{s \theta}$, this can be written as 
\begin{gather}
Q_{ss'}^\text{fric} = \frac{m_{s'}}{m_s + m_{s'}} \big( v_{s'\theta} - v_{s \theta} \big) R_{ss'\theta} , 
\end{gather}
where the friction force density $R_{ss'\theta}$ is 
\begin{align}
&R_{ss'\theta} = n_s m_s \nu_{ss'} ( v_{s' \theta} - v_{s \theta} ) \nonumber \\
&\hspace{40 pt}+ \frac{n_s m_s \nu_{ss'}}{B} \bigg( \frac{\gamma_{ss'} T_s T_{s'}}{q_s} - \frac{\gamma_{s's} T_{s'} T_{s'}'}{q_{s'}} \bigg) , 
\end{align}
where 
\begin{gather}
\gamma_{ss'} \doteq \frac{3}{2} \frac{m_{s'}}{m_s T_{s'} + m_{s'} T_s}. 
\end{gather}
The momentum equation for species $s$ is given by Eq.~(\ref{eqn:momentum}). In steady state, its $\hat r$ component is 
\begin{align}
&m_s n_s v_{sr} \frac{\partial v_{sr}}{\partial r} - m_s n_s \frac{v_{s \theta}^2}{r} \nonumber \\
&\hspace{50 pt}= q_s n_s E + q_s n_s v_{s \theta} B -  \frac{\partial p_s}{\partial r} + R_{sr}. 
\end{align}
$v_{sr}$ is small compared to $v_{s \theta}$, so $m_s n_s v_{sr} \partial v_{sr} / \partial r \ll m_s n_s v_{s \theta}^2 / r$. The collision frequencies $\nu_{ss'}$ are small compared to the gyrofrequency $\Omega_s$, so $R_{sr} \ll q_s n_s v_{s \theta} B$. Dropping these two small terms and rearranging, 
\begin{gather}
v_{s \theta} = - \frac{E}{B} + \frac{1}{q_s B n_s} \frac{\partial p_s}{\partial r} - \frac{1}{\Omega_s} \frac{v_{s \theta}^2}{r} \, .
\end{gather}
Then 
\begin{align}
&v_{s' \theta} - v_{s \theta} = \nonumber \\
&\hspace{30 pt}\frac{1}{q_s B} \bigg[ \frac{q_s}{q_{s'}} \frac{p_{s'}'}{n_{s'}} - \frac{p_s'}{n_s} - \frac{q_s}{q_{s'}} \frac{m_{s'} v_{s' \theta}^2}{r} + \frac{m_s v_{s \theta}^2}{r} \bigg] 
\end{align}
and 
\begin{align}
&R_{ss' \theta} = \frac{n_s m_s \nu_{ss'}}{q_s B} \bigg[ \frac{q_s}{q_{s'}} \frac{p_{s'}'}{n_{s'}} - \frac{p_s'}{n_s} - \frac{q_s}{q_{s'}} \frac{m_{s'} v_{s' \theta}^2}{r} + \frac{m_s v_{s \theta}^2}{r} \nonumber \\
&\hspace{75 pt}+ \bigg( \gamma_{ss'} T_s T_{s}' - \frac{q_s}{q_{s'}} \gamma_{s's} T_{s'} T_{s'}' \bigg) \bigg] .
\end{align}
The condition for $R_{ss'\theta} = 0$ can be rewritten as 
\begin{align}
&\frac{1}{q_{s'}} \bigg( \frac{p_{s'}'}{n_{s'}} - \frac{m_{s'} v_{s' \theta}^2}{r} - \gamma_{s's} T_{s'} T_{s'}' \bigg) \nonumber \\
&\hspace{70 pt}= \frac{1}{q_s} \bigg( \frac{p_s'}{n_s} - \frac{m_s v_{s \theta}^2}{r} - \gamma_{ss'} T_s T_s' \bigg) . 
\end{align}
In the simple limit where $T_{s'}(r) = \tau T_{s}(r)$ for some constant $\tau$, 
\begin{align}
&\bigg\{ \frac{p_s(r)}{p_{s}(0)} \exp \bigg[- \int_0^r \D r \bigg( \frac{m_s v_{s \theta}^2}{r T_s} + \gamma_{ss'} T_s'
\bigg) \bigg] \bigg\}^{1/Z_s} \nonumber \\
&\hspace{10 pt}=\bigg\{ \frac{p_{s'}(r)}{p_{s'}(0)} \exp \bigg[ - \int_0^r \D r \bigg( \frac{m_{s'} v_{s' \theta}^2}{r T_{s'}} + \gamma_{s's} T_{s'}' \bigg) \bigg] \bigg\}^{\tau / Z_{s'}} . \label{eqn:pFirstKind}
\end{align}
where $Z_s \doteq q_s / e$, so electrons would have $Z_e = -1$. The condition for $v_{s' \theta} - v_{s \theta} = 0$ can similarly be treated similarly. Under the same assumptions, it reduces to 
\begin{align}
&\bigg\{ \frac{p_s(r)}{p_{s}(0)} \exp \bigg[- \int_0^r \D r \, \frac{m_s v_{s \theta}^2}{r T_s} \bigg] \bigg\}^{1/Z_s} \nonumber \\
&\hspace{30 pt}=\bigg\{ \frac{p_{s'}(r)}{p_{s'}(0)} \exp \bigg[ - \int_0^r \D r \, \frac{m_{s'} v_{s' \theta}^2}{r T_{s'}} \bigg] \bigg\}^{\tau / Z_{s'}} . \label{eqn:pSecondKind}
\end{align}
The condition for $R_{ss'\theta} = 0$ and the condition for $v_{s \theta} = v_{s'\theta}$ are identical in the limit where $T_s' = T_{s'}' = 0$. 

Define $P^R_{ss'}$ as a profile for species $s$ that satisfies Eq.~(\ref{eqn:pFirstKind}) with respect to species $s'$:
\begin{align}
&P^R_{ss'}(r) = P^R_{ss'}(0) \exp \bigg[\int_0^r \D r \bigg( \frac{m_s v_{s \theta}^2}{r T_s} + \gamma_{ss'} T_s'
\bigg) \bigg] \nonumber \\
&\hspace{0 pt}\times\bigg\{ \frac{p_{s'}(r)}{p_{s'}(0)} \exp \bigg[ - \int_0^r \D r \bigg( \frac{m_{s'} v_{s' \theta}^2}{r T_{s'}} + \gamma_{s's} T_{s'}' \bigg) \bigg] \bigg\}^{\tau Z_s / Z_{s'}} . 
\end{align}
Define $P^v_{ss'}$ as a profile for species $s$ that satisfies Eq.~(\ref{eqn:pSecondKind}) with respect to species $s'$: 
\begin{align}
& P^v_{ss'}(r) = P^v_{ss'}(0) \exp \bigg[\int_0^r \D r \, \frac{m_s v_{s \theta}^2}{r T_s} \bigg] \nonumber \\
&\hspace{10 pt}\times \bigg\{ \frac{p_{s'}(r)}{p_{s'}(0)} \exp \bigg[ - \int_0^r \D r \, \frac{m_{s'} v_{s' \theta}^2}{r T_{s'}} \bigg] \bigg\}^{\tau Z_s / Z_{s'}} . 
\end{align}
The flux $\Gamma_{ss'}^\text{fric}$ tends to make $p_s$ relax to $P_{ss'}^R$. Note that the electric field appears only in the centrifugal $v_{s\theta}^2$ term (unlike in the unmagnetized case, where it can drive differential transport more directly \cite{Kagan2012, Kagan2014ii}). 

The heating $Q_{ss'}^\text{fric}$ can be rewritten in a way that more explicitly shows how it depends on the deviation of $p_s$ from $P_{ss'}^R$ and $P_{ss'}^v$:
\begin{align}
&Q_{ss'}^\text{fric} = \frac{m_{s'}}{m_s + m_{s'}} \frac{T_s^2 n_s m_s \nu_{ss'}}{q_s^2 B^2} \nonumber \\
&\hspace{30 pt}\times \bigg[ \frac{\partial}{\partial r} \log \bigg( \frac{P_{ss'}^R}{p_s} \bigg) \bigg] \bigg[ \frac{\partial}{\partial r} \log \bigg( \frac{P_{ss'}^v}{p_s} \bigg) \bigg] .
\end{align}
This expression vanishes as $p_s$ approaches either $P_{ss'}^R$ or $P_{ss'}^v$. 

\section{Compressional Heating} \label{appendix:compressionalHeating}

The temperature evolution equation, Eq.~(\ref{eqn:heat}), includes a compressional heating term that is largely not discussed in this paper:
\begin{gather}
\frac{3 n_s}{2} \frac{dT_s}{dt} \bigg|_\text{compressional} = - p_s \nabla \cdot \bv_s . 
\end{gather} 
In part, this is a question of scope: the focus of the paper is on the classical heating terms $Q_s^\text{visc}$ and $Q_s^\text{fric}$, and on their connection with the classical particle fluxes $\Gamma_s^\text{visc}$ and $\Gamma_s^\text{fric}$. 
Moreover, the compressional heating behaves identically in a rotating plasma as in any other plasma (unlike $Q_s^\text{visc}$ and $Q_s^\text{fric}$, both of which depend on the rotation profile in one way or another). 

However, it may be helpful to say something about the expected size of the compressional heating. If $s_s$ is the volumetric particle source rate, then the continuity equation is 
\begin{gather}
\frac{\partial n}{\partial t} + \nabla \cdot (n_s \bv_s) = s_s. 
\end{gather}
Then for a system in steady state, 
\begin{align}
-p_s \nabla \cdot \bv_s &= - T_s \nabla \cdot (n_s \bv_s) + T_s \bv_s \cdot \nabla n_s \\
&= -T_s s_s + T_s \bv_s \cdot \nabla n_s. 
\end{align}
Note that for inflowing particles, $\bv_s \cdot \nabla n_s$ is positive if the density is peaked toward the core of the system. 
Away from sources, if the density has a gradient scale length $L$, 
\begin{align}
-p_s \nabla \cdot \bv_s \sim \frac{T_s \Gamma_s}{L} \, . \label{eqn:compressionalScaling}
\end{align}
This can be contrasted with Eq.~(\ref{eqn:integratedViscousHeating}), which implies that without boundary stresses, 
\begin{gather}
Q_s^\text{visc} \sim q_s E \Gamma_s^\text{visc}. 
\end{gather}
For many rapidly rotating systems, $q_s E$ is large compared to $T_s / L$. In the most dramatic cases discussed in this paper, where the $\mathbf{j} \cdot \bE$ heating is very large and can be directed into the ions, the scaling in Eq.~(\ref{eqn:compressionalScaling}) suggests that the compressional heating will not be important. However, that does not mean there are no scenarios in which it matters; a full solution of the temperature evolution equation would have to take it into account. 

\end{appendix}

	\bibliography{../../../../Master.bib}
	
\end{document}